\documentclass[12pt,aps,prd,superscriptaddress,showpacs,longbibliography,floatfix,nofootinbib]{revtex4-2}

\usepackage[utf8]{inputenc}
\usepackage{slashed}
\pdfoutput=1

\usepackage{color}
\usepackage{graphicx}   
\usepackage{bm}
\usepackage{amsmath}
\usepackage{amsfonts}
\usepackage{eufrak}
\usepackage{hyperref}

\newcommand{\be}{\begin{equation}}
\newcommand{\ee}{\end{equation}}
\newcommand{\ba}{\begin{eqnarray}}
\newcommand{\ea}{\end{eqnarray}}

\begin{document}

\title{Variational Solution of Non-Hermitian Quantum Field Theory in Two Dimensions}

\author{Paul Romatschke}
\affiliation{Institute für Theoretische Physik, TU Wien, Wiedner Hauptstraße 8-10, 1040 Wien, Austria}

\begin{abstract}
  This work describes a variational technique to solve non-Hermitian quantum systems, in particular those with negative coupling quartic self-interaction. Differences of applying variational methods from standard quantum mechanics to non-Hermitian quantum mechanics -- including potential pitfalls --  are highlighted. A successful implementation of the variational approach for the non-Hermitian case is presented. The technique is used to analyze scalar quantum field theory in d=2 with negative quartic self-interaction in the Hamiltonian formulation in lattice discretization.
\end{abstract}

\maketitle

\section{Introduction}

The variational method is a versatile tool to obtain the Hamiltonian eigenspectrum in quantum mechanics. In ordinary quantum mechanics, where the Hamiltonian is Hermitian, ${\cal H}^\dagger={\cal H}$, the variational principle guarantees that choosing a finite set of basis vectors will result in a strict upper bound for the ground state energy $E_0$ of the system \cite{Yuan_2019}. In addition, for a sufficiently large number of basis states, the variational method can be extended to quantum field theories in the Hamiltonian formulation, cf. Ref.~\cite{Romatschke:2026qxm}.

In this work, I consider applying the variational method to non-Hermitian field theories, in particular field theories which are invariant under the combined action of parity flip ${\cal P}$ and time reversal ${\cal T}$, so called $\cal PT$-symmetric quantum field theories. Since the pioneering work by Bender and B\"ottcher on ${\cal PT}$-symmetric quantum mechanics \cite{Bender:1998ke}, interest in the field of non-Hermitian systems has grown tremendously, cf. Refs.~ \cite{2010NatPh...6..192R,2018NatPh..14...11E}, including applications to quantum field theory \cite{PhysRevLett.40.1610,PhysRevLett.54.1354,Li:2024xms,ArguelloCruz:2025zuq,Bender:2018pbv,Felski:2021evi,Branchina:2021czr,Beygi:2019qab,Felski:2020vrm,Mavromatos:2020hfy,Felski:2021bdg,Mavromatos:2021hpe,Ai:2022csx,Romatschke:2022jqg,Lawrence:2023woz,Weller:2023jhc,Chen:2024ynx,Romatschke:2024cld,Barberena:2025ibo,Fring:2019hue,Fring:2020bvr,Romatschke:2024hpb}, finite-density QCD on the lattice \cite{Ogilvie:2024vde} and gravity \cite{Mavromatos:2024ozk,Kuntz:2024rzu}.

A practical issue occurring for non-Hermitian field theories is the accurate study of their non-perturbative properties. While traditional expansion methods work well whenever there is a small parameter to expand in \cite{Bender:2018pbv,Romatschke:2022jqg,Grable:2023paf}, the application of Monte Carlo sampling techniques for lattice implementations of non-Hermitian field theories is hindered by the presence of a strong sign problem, cf. Ref.~\cite{Romatschke:2023fax}, requiring new ideas such as adaptive contour deformations \cite{Lawrence:2022afv} or complex Langevin techniques \cite{Aarts:2017hqp}. A promising alternative to lattice methods for non-Hermitian field theories could also be Hamiltonian truncation techniques, such as those explored in Ref.~\cite{Lencses:2024wib}.

This work is a generalization of the variational method as implemented in Ref.~\cite{Romatschke:2026qxm} to the case of non-Hermitian field theories, specifically field theories with negative quartic self interaction, which is of particular relevance for the physics of the Higgs boson in the Standard Model of Particle Physics \cite{Symanzik:1973hx,Romatschke:2023sce,Romatschke:2023ogd}. Because these theories are non-Hermitian, many properties of the variational principle for Hermitian systems are no longer valid. For this reason, the method is first tested for the case of non-Hermitian quantum mechanics where an isospectral formulation to a Hermitian Hamiltonian is known \cite{Jones:2006qs}, see section \ref{sec:qm}. Using lessons learned from this test case, the method is then applied to non-Hermitian scalar quantum field theory in section \ref{sec:qft}, after which a summary of results and conclusions is provided in section \ref{sec:conc}.

\section{``Wrong-Sign'' Quartic Oscillator in Quantum Mechanics}
\label{sec:qm}

As a start, let me discuss an application of the variational method to a well-studied non-Hermitian system, namely the ``wrong-sign'' quartic oscillator with Hamiltonian in the position basis
\be
\label{hamiltonian}
{\cal H}=-\frac{1}{2}\frac{d^2}{dx^2}+\frac{m_B^2x^2}{2}-g x^4\,.
\ee
Here $g\geq 0$ is the coupling constant of the theory, which carries mass dimension 3, and $m_B^2\in \mathbb{R}$ is the mass squared which can take on any real value. Choosing a complete set of states $|n\rangle$, one can construct the matrix elements
\be
\label{matrixelements}
H_{nm}\equiv \langle n| {\cal H}|m\rangle=\int dx \langle n|x\rangle \langle x| {\cal H}|m\rangle\,,
\ee
where
\be
\langle x| n\rangle\equiv \psi_n(x)\,,
\ee
are the state-dependent wave-functions in the position basis. In the following, I will consider non-Hermitian systems with ${\cal PT}$ symmetry, and as a consequence it is useful to employ basis functions which are also ${\cal PT}$-symmetric, or
\be
\label{symfunc}
\psi_n^{\cal PT}(x)=\psi_n^*(-x)=\psi_n(x)\,.
\ee

A crucial difference with respect to ordinary Hermitian quantum mechanics is that the wave-functions $\langle n |x\rangle$ are \textbf{not} the complex conjugate of $\psi_n(x)$. For systems that are non-Hermitian, but ${\cal PT}$-symmetric, the wave-functions $\langle n |x\rangle$ are instead the ${\cal PT}$ conjugate of the wave-functions $\psi_n(x)$:
\be
\langle n |x\rangle=\left(\psi_n(x)\right)^{\cal PT}=\psi_n^*(-x)=\psi_n(x)\,,
\ee
where I have used (\ref{symfunc}), see the discussion in sec. VIB of Ref.~\cite{Bender:2023cem}.

Another difference with respect to Hermitian quantum mechanics is that the inner-product for the wave-functions is
\be
\label{funnynorm}
\int_{\cal C} dx \psi_{n}(x)\psi_m(x)=(-1)^n \delta_{nm}\,,
\ee
where ${\cal C}$ is an appropriate contour in the complex plane, and the factor $(-1)^n$ is required by the ${\cal PT}$-symmetric inner product \cite{Bender:2023cem}. In practice, I choose the ``triangle'' contour ${\cal C}$ for the integration \cite{Lawrence:2023woz}, which is parametrized as
\be
\label{trianglecontour}
x=s \left(e^{-i\alpha}\Theta(s)+e^{i \alpha}\Theta(-s)\right)\,,\quad s\in \mathbb{R}\,,
\ee
such that (\ref{funnynorm}) becomes
\be
\label{orthoG}
(-1)^n \delta_{nm} = \int_0^\infty ds \left[e^{-i\alpha} \psi_n(s e^{-i\alpha})\psi_m(s e^{-i\alpha})+e^{i\alpha} \psi_n(-s e^{i\alpha})\psi_m(-s e^{i\alpha})\right]\,.
\ee

For the case of the ``wrong-sign'' quartic oscillator with Hamiltonian (\ref{hamiltonian}), I choose $\alpha=\frac{\pi}{6}$ and construct trial wave-functions
\be
\psi_0(x)=n_0 e^{\frac{w}{3}(i x)^3},\quad \psi_1(x)=n_1 e^{\frac{w}{3}(i x)^3} \left(i x+c_{10}\right)\,,\quad \psi_2(x)=n_2 e^{\frac{w}{3}(i x)^3} \left((i x)^2+c_{21}(ix)+c_{20}\right)\,,
\ee
where $n_0,n_1,n_2$ are normalization factors and $w$ is a variational parameter. Both the coefficients $c_{10},c_{21},c_{20}$ and the normalization factors are determined by requiring (\ref{orthoG}) to be fulfilled for any pair of wave-functions. The choice of $\alpha$ implies
\ba
\psi_0(\pm s e^{\mp i\alpha})&=&n_0 e^{-\frac{w}{3} s^3}\,,\nonumber\\
\psi_1(\pm s e^{\mp i\alpha})&=&n_1 e^{-\frac{w}{3} s^3}(s e^{\pm \frac{i\pi}{3}}+c_{10})\,,\nonumber\\
\psi_2(\pm s e^{\mp i\alpha})&=&n_2 e^{-\frac{w}{3} s^3}(s^2 e^{\pm \frac{2i\pi}{3}}+c_{21}s e^{\pm \frac{i\pi}{3}}+c_{20})\,,
\ea
where the upper (lower) sign corresponds to the evaluation of the functions $\psi_n(x)$ on the right (left) part of the contour ${\cal C}$. This informs the parametrization of the functions $\psi_n(x)$ as
\be
\label{negcouppsin}
\psi_n\left(\pm s e^{\mp i\alpha}\right)=n_n e^{\frac{w z^3}{3}}P_n(z)\equiv \chi_n(z)\,,\quad z=s e^{\pm \frac{i\pi}{3}}\,,
\ee
where $P_n(z)$ are real-valued polynomials of argument $z$. One finds
\ba
\label{negwaffis}
P_0(z)&=&1, \quad n_0^{-2}=\frac{\Gamma\left(\frac{1}{3}\right)}{2^{\frac{1}{3}}3^{\frac{1}{6}} w^{\frac{1}{3}}}\,,\\
P_1(z)&=&z-\frac{\sqrt{\pi}}{\Gamma\left(\frac{1}{6}\right)}\left(\frac{6}{w}\right)^{\frac{1}{3}}\,, \quad n_1^{-2}=\frac{2^{\frac{1}{3}}\sqrt{3}\pi\Gamma\left(\frac{1}{3}\right)}{w \Gamma\left(\frac{1}{6}\right)^2}\,,\nonumber\\
P_2(z)&=&z^2-\left(\frac{6}{w}\right)^{\frac{1}{3}}\frac{3 \Gamma^2\left(\frac{7}{6}\right)}{\pi} P_1(z)\,,\quad n_2^{-2}\simeq \frac{0.297631}{w^{\frac{5}{3}}}\nonumber\,,\\
P_3(z)&\simeq &z^3-\frac{2.57022}{w^{\frac{1}{3}}}P_2-\frac{0.86413}{w^{\frac{2}{3}}}P_1+\frac{1}{2w}\,,\quad n_3^{-2}\simeq\frac{0.19566}{w^{\frac{7}{3}}}\,,\nonumber
\ea
and higher order polynomials are readily generated  from (\ref{orthoG}). Specifically, parametrizing
\be
P_n(z)=z^n+\sum_{i=1}^n c_{n,i}P_{n-i}(z)\,,
\ee
Eq.~(\ref{orthoG}) implies for the coefficients $c_{n,i}$ and normalizations $n_n$
\be
c_{n,i}=-\frac{\langle z^{n}P_{n-i}(z)\rangle}{n_{n-i}^2}\,,\quad
n_n^{-2}=(-1)^n\left[\langle z^{2n}\rangle-\sum_{i=1}^n c_{n,i}^2 n_{n-i}^2 \right]\,,
\ee
where $\langle\cdot \rangle$ denotes the expectation value on the contour ${\cal C}$ with appropriate weights.

\subsection{Massless case $m=0$}

To introduce the method and discuss its limitations, let me first discuss the massless case $m=0$ of the Hamiltonian (\ref{hamiltonian}).

With the basis functions known, the matrix elements (\ref{matrixelements}) are readily evaluated numerically for any value of $w$. Truncating the matrix $H_{nm}$ at $n\leq K+1$ one can calculate eigenvalues $e_n$ for the truncated matrix
\be
H_{nm}\chi_m^{(n)}=e_n \chi_m^{(n)}\,,\quad H^\dagger_{nm}\tilde \chi_m^{(n)}=e_n^* \tilde\chi_m^{(n)}\,,
\ee
where $\chi^{(n)},\tilde \chi^{(n)}$ are the right and left eigenstates of the matrix $H_{nm}$ and $e_n$ the corresponding eigenvalues. In the limit $K\rightarrow \infty$, the eigenvalues $e_n$ of the truncated matrix will correspond to the eigenvalues $E_n$ of the Hamiltonian (\ref{hamiltonian}),
\be
\lim_{K\rightarrow \infty} e_n=E_n\,,
\ee
and for unbroken ${\cal PT}$ symmetry one has
\be
\lim_{K\rightarrow \infty}\tilde \chi^{(n)}_m=(-1)^m  \chi^{(n)}_m\,.
\ee

Note that expectation values for a generic operator ${\cal O}$ in the ground state of the theory can be calculated as
\be
\langle {\cal O}\rangle = \frac{\langle \tilde \chi^{(0)}|{\cal O}|\chi^{(0)}\rangle}{\langle \tilde \chi^{(0)}|\chi^{(0)}\rangle}\,,
\ee
cf. Ref.~\cite{3kk2-3fsj}.

For finite truncation $K$, $e_n$ will still depend on the variational parameter $w$. For instance, for $K=1$ one has
\be
H_{00}=\langle 0 |{\cal H}|0\rangle=\left(\frac{3}{2w^4}\right)^{\frac{1}{3}}\frac{3(2g+w^2)\Gamma\left(\frac{5}{3}\right)}{4\Gamma\left(\frac{1}{3}\right)}\,,
\ee
and the variational minimum for this matrix element is located at $w=2\sqrt{g}$, so that
\be
K=1:\quad e_0=\frac{9 (3 g)^{\frac{1}{3}} \Gamma\left(\frac{5}{3}\right)}{4 2^{\frac{2}{3}} \Gamma\left(\frac{1}{3}\right)}\simeq 0.6889 g^{\frac{1}{3}}\ldots\,.
\ee

\begin{figure}[t]
  \includegraphics[width=.8\linewidth]{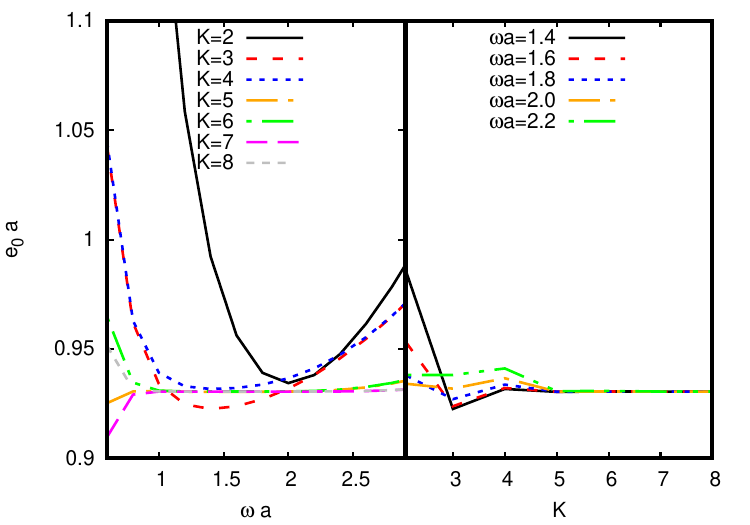}
  \caption{\label{fig:merged} Lowest-lying eigenvalue $e_0$ of the Hamiltonian matrix $H_{nm}$ truncated at $n=K+1$ for variational parameter $\omega$. Left side shows dependence for fixed truncation $K$ on varying $\omega$, right side shows dependence for fixed $\omega$ on varying $K$. See text for details.}
\end{figure}

Including more  wave-functions and repeating the variational procedure, one finds that there are multiple local minima. Moreover, for certain K, $e_0$ is not bounded from below as a function of $w$, see Fig.~\ref{fig:merged} This situation is expected: unlike Hermitian matrices, where the variational formulation is guaranteed to possess a local minimum for finite $K$ which bounds the true lowest energy eigenvalue from above, no such theorem exists for non-Hermitian (${\cal PT}$-symmetric) Hamiltonians and finite number of truncation level $K$. 

However, for ${\cal PT}$-symmetric Hamiltonians one can use another property from the variational approach that transfers over from Hermitian systems: the lowest energy eigenvalue of ${\cal H}$ for given truncation K is located at the minimum with respect to the variation of the truncation level K, see Fig.~\ref{fig:merged}. This is implemented as follows: for given truncation level K, I calculate
\be
\Delta_K e_0=(e_0^{(K)}-e_0^{(K-1)})^2
\ee
as a function of $w$ and report $e_0^{(K)}$ determined by the value of $w$ where $\Delta_K$ attains its minimum. Defining $\hat e_0=\frac{e_0}{g^{\frac{1}{3}}}$ leads to the results shown in Tab.~\ref{tab:one}

\begin{table}
  \begin{tabular}{|c|ccccccc|}
    \hline
    K   & 2 & 3 & 4 & 5 & 6 & 7 & 8\\
    \hline
    $\hat{e}_0$  & 0.934399  & 0.938095 & 0.976276  & 0.93038 & 0.930802 & 0.930539 &0.930546\\
    \hline
    rel. err. &  $4\times 10^{-3}$ & $8\times 10^{-3}$ & $5\times 10^{-2}$ 
    & $-2\times 10^{-4}$ & $3\times 10^{-4}$ & $-8\times 10^{-6}$ & $1\times 10^{-6}$\\
    \hline
  \end{tabular}
  \caption{\label{tab:one} Lowest energy eigenstate of the Hamiltonian (\ref{hamiltonian}) for $m_B=0$ using the variational method. See text for details.}
  \end{table}

In this table, the relative errors have been calculated using the ground state energy result $\hat{E}_0\simeq 0.93054606\ldots$ from Refs.~\cite{Bender:1998ke,Romatschke:2024mxr}.   As can be seen from this table, the variational method is capable of obtaining the ground-state energy also for the negative-coupling theory. However, the approach to the ground state energy $\hat E_0$ as $K$ is increased is not monotonous, unlike the case for the positive coupling theory, cf. Fig.~\ref{fig:merged}.

\begin{figure}[t]
  \includegraphics[width=.8\linewidth]{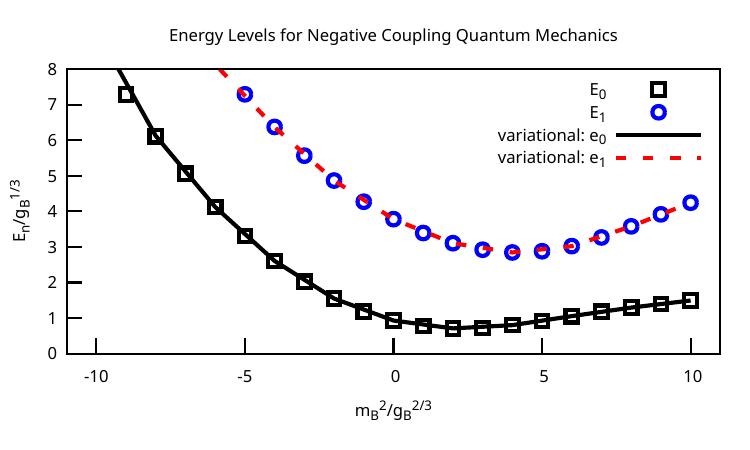}
  \caption{\label{fig:one} Energy levels of the negative coupling Hamiltonian (\ref{hamiltonian}) with $m_B^2\neq 0$ calculated using different techniques and approximations. Shown are results from the variational method for the lowest lying eigenvalue $e_0$ and first excited state $e_1$ (lines). For comparison, results for the ground state and first excited state energies $E_0,E_1$ obtained using direct numerical diagonalization of the isospectral Hamiltonian (squares, circles) are shown \cite{Jones:2006qs,Bender:2007nj}. See text for details.}
\end{figure}

\subsection{Massive Case}

I have implemented the above variational procedure to find the two lowest-lying eigenvalues $e_0,e_1$ for the massive Hamiltonian (\ref{hamiltonian}), with results shown in Fig. \ref{fig:one}. These results are compared to the ground state energy and first excited state energy from exact diagonalization of the isospectral Hamiltonian ${\cal H}_{\rm iso}=\frac{p^2}{2}+4 g x^4+\sqrt{2 g x}-x^2 m_B^2+\frac{m_B^4}{16g}$, cf. Refs.~ \cite{Jones:2006qs,Bender:2007nj}.

As can be seen from this Figure, there is broad overall agreement for both energy eigenvalues of the Hamiltonian (\ref{hamiltonian}) between the different methods. In particular, the numerical agreement between the variational method and the eigenvalues of the isospectral Hamiltonian ${\cal H}_{\rm iso}$ suggests that the variational method as outlined above works and converges to the correct answer for negative coupling systems.

  \section{Negative Coupling $\phi^4$ theory in Two Dimensions}
\label{sec:qft}
  Let me now apply the techniques tested in the previous sections to the case of $\phi^4$ field theory in two Euclidean dimensions with negative coupling $\lambda=-g$, defined by the Euclidean action
  \be
  \label{mainaction}
S=\int d^2x \left[\frac{1}{2}\partial_\mu \phi \partial_\mu \phi+\frac{m_B^2 \phi^2}{2}-g\phi^4\right]\,.
  \ee

This theory is studied on a lattice by discretizing the action on $D$ points along the ``space'' direction with length $L$, such that
\be
\label{latticeaction}
S_{(D)}=a \int d\tau \sum_{k=0}^{D-1} \left[\frac{\dot{\phi}^2(\tau,x_k)}{2}+\frac{\left(\phi(\tau,x_{k+1})-\phi(\tau,x_k)\right)^2}{2a^2}+\frac{m_B^2 \phi^2(\tau,x_k)}{2}-g \phi^4\left(\tau,x_k\right)\right]\,,
\ee
where $a=\frac{L}{D}$ is the lattice spacing, $x_k=-\frac{L}{2}+k a$ with $k\in \mathbb{N}$ is the location on the discretized spatial direction, and $\dot{\phi}=\frac{d\phi}{d \tau}$ is the (continuous) derivative along the ``time''  direction and $\lim_{D\rightarrow \infty} S_{(D)}=S$. Following the setup outlined for the positive coupling theory in Ref.~\cite{Romatschke:2026qxm}, I scale $\phi\rightarrow \frac{\phi}{\sqrt{a}}$ and use the notation
\be
\phi(\tau,x_k)=\phi_k(\tau)\equiv x_k\,.
\ee
In this fashion, the problem turns into quantum mechanics in $D$ dimensions with Hamiltonian
  \be
  {\cal H}=\frac{\vec{p}^2}{2}+V(\vec{x})\,,\quad V(\vec{x})=\frac{m_B^2 \vec{x}^2}{2}+\sum_{k=0}^{D-1}\left[\frac{(x_{k+1}-x_k)^2}{2 a^2}-\frac{g}{a} x_k^4\right]\,.
  \ee

  I construct a basis for the quantum mechanics problem as the product states of the one-dimensional negative coupling theory, e.g. products of the wave-functions $\psi_n$ given in (\ref{negcouppsin}),
  \be
  \label{basiselement}
  \langle \vec{x}|n_1 n_2\ldots n_{D}\rangle = \psi_{n_1}(x_1)\psi_{n_2}(x_2)\ldots \psi_{n_D}(x_D)\,.
  \ee
  The matrix elements are calculated by performing the integration over $x$ on a complex contour ${\cal C}$, which for simplicity is again taken to be the triangle contour (\ref{trianglecontour}). Particular care must be taken to correctly include the normalization factors (\ref{funnynorm}).

  As an example, let's calculate the matrix element for the state
  \be
  \langle \vec{x}| 00\ldots 0 \rangle = \prod_{k=1}^{D}\psi_0(x_k)\,.
  \ee
  One has
  \ba
  H_{00}=\int_{\cal C} d\vec{x} \langle 00\ldots 0|\vec{x}\rangle \langle {\cal H} \vec{x} |00\ldots 0\rangle
  &=&D\int_{\cal C} dx \psi_0(x)\left[-\frac{1}{2}\frac{d^2}{dx^2}+\left(\frac{m_B^2}{2}+\frac{1}{a^2}\right)x^2 -\frac{g}{a}x^4\right]\psi_0(x)\nonumber\\
  &&-\frac{D}{a^2}\left(\int_{\cal C} dx \psi_0(x) x \psi_0(x)\right)^2\,.
  \ea

  Using the contour-variable relation $x=-i z$ with $z=s e^{\pm \frac{i \pi}{3}}$ from (\ref{negcouppsin}) and $s\in \mathbb{R}^+$, one has
  \ba
  H_{00}&=&D\int_0^\infty ds e^{-i\frac{\pi}{6}}\chi_0(z) \left[\frac{1}{2}\frac{d^2}{dz^2}-\left(\frac{m_B^2}{2}+\frac{1}{a^2}\right)z^2-\frac{g}{a}z^4\right]\left.\chi_0(z)\right|_{z=s e^{\frac{i\pi}{3}}}+{\rm cc.}
  \nonumber\\
  &&  +\frac{D}{a^2}\left(\int_0^\infty ds e^{-i\frac{\pi}{6}}\chi_0(z) z \left.\chi_0(z)\right|_{z=s e^{\frac{i\pi}{3}}}+{\rm cc.}\right)^2\,,
  \ea
  where ${\rm cc.}$ denotes complex conjugation. Given the explicit form of $\chi_0(z)=n_0 e^{\frac{w z^3}{3}}$ from the previous section, one obtains for the matrix element
  \be
  H_{00}=Dn_0^2\left(\frac{3^{\frac{1}{6}}\Gamma\left(\frac{2}{3}\right)\left(w^2+\frac{2 g}{a}\right)}{2^{\frac{5}{3}}w^{\frac{5}{3}}}+n_0^2\frac{3^{\frac{1}{3}}\Gamma^2\left(\frac{2}{3}\right)}{2^{\frac{4}{3}} a^2 w^{\frac{4}{3}}}\right)\,,
  \ee
  with the normalization $n_0$ given in (\ref{negwaffis}).

\begin{figure}[t]
  \includegraphics[width=.8\linewidth]{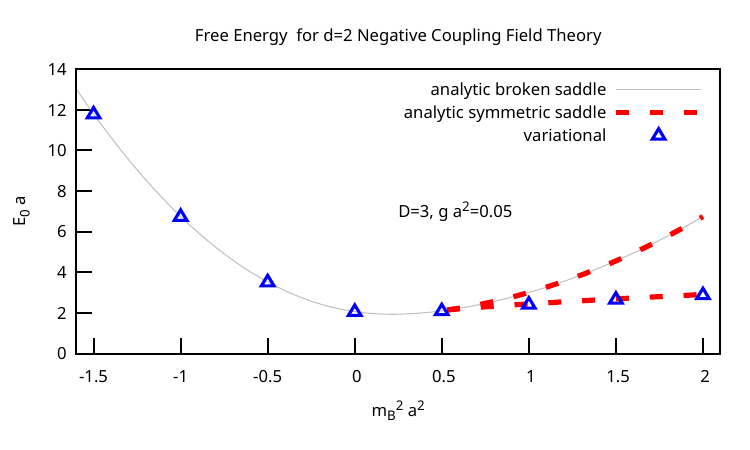}
  \caption{\label{fig:two} Free energy of negative coupling QFT on a lattice with $D=3$ transverse sites and coupling $g a^2=0.05$. Shown are results from the variational method (symbols) as well as from the analytic saddle-point expansion \cite{Romatschke:2026zvd} (lines). See text for details.}
\end{figure}
  
  Other matrix elements for fixed number of transverse sites $D$ are calculated in a similar manner, so that one ends up with a large matrix $M_H$ with elements
  \be
  \label{QFTmatrixelements}
I_{n_1n_2\ldots n_D}^{m_1 m_2\ldots m_D}\equiv \langle n_1 n_2\ldots n_D|{\cal H}|m_1 m_2\ldots m_D\rangle\,,
\ee
 that can be diagonalized numerically. Note that $M_H$ is not Hermitian since the theory under consideration is not Hermitian. As a consequence, the eigenvalue spectrum of $M_H$ is in general complex.

 For a set of $K$ basis states, I calculate the eigenvalue spectrum of $M_H$ for fixed values of $D, m_B^2 a^2, g a^2$ and $w a^{\frac{3}{2}}$. Eigenvalues $e_n$ of $M_H$ are sorted according to the magnitude of their respective real parts for a range of values of $w a^{\frac{3}{2}}$. Increasing $K$ by a fixed increment $\Delta K$, (e.g. $\Delta_K=100$ for $D=3$, $\Delta_K=1000$ for $D=5$), I calculate
 \be
 \Delta_K {\rm Re}(e_0)=[{\rm Re}(e_0^{(K)}-e_0^{(K-\Delta_K)})]^2
 \ee
 as a function of $w a^{\frac{3}{2}}$ and report $e_0^{(K)}$ determined by the value of $w$ where $\Delta_K$ attains its minimum. I find that the resulting value of $e_0^{(K)}$ using this procedure is real within machine precision.

 As an example, results for $e_0^{(K)}$ for $D=3$ and fixed coupling strength $g a^2=0.05$ as a function of $m_B^2 a^2$ are shown in Figure \ref{fig:two}. For guidance, results from an (approximate) analytic saddle-point method from Ref.~\cite{Romatschke:2026zvd}, adapted to the lattice theory for the same parameters   are also shown in Fig. ~\ref{fig:two}, see Eq.~(\ref{relationSP}). As can be seen from this figure, the agreement between the variational method and the (approximate) analytic saddle-point method is quantitatively good, as was observed for the positive coupling theory \cite{Romatschke:2026qxm}.

I find that the energy spectrum $e_n^{(K)}$ calculated at fixed $K$ typically contains complex conjugate pairs. However, as the number of basis states $K$ is increased, the eigenvalues $e_n$ with the smallest real part stabilize and their imaginary part decreases, until all low-lying $e_n^{(K)}$ are purely real.

\begin{figure}[t]
  \includegraphics[width=.8\linewidth]{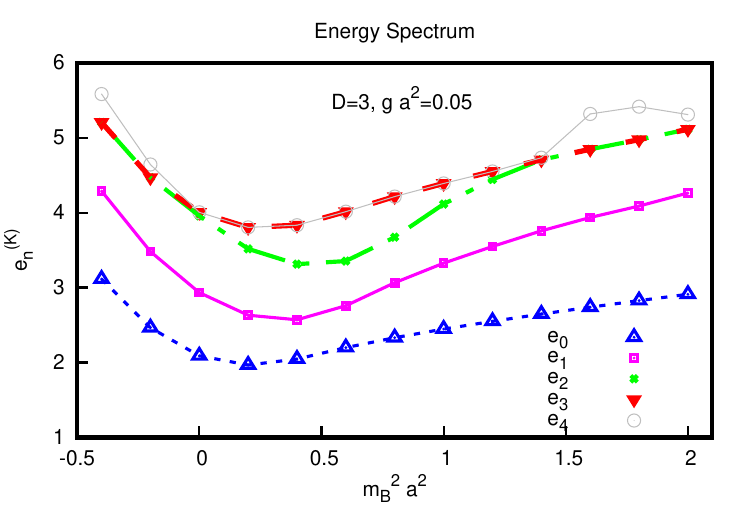}
  \caption{\label{fig:mdep} Low-lying eigenvalues Hamiltonian eigenvalues $e_n^{(K)}$ in the variational method truncated at $K=1000$ basis states for negative coupling $\phi^4$ theory as a function of $m_B^2 a^2$. See text for details.}
\end{figure}

An example for the low-lying Hamiltonian eigenvalue spectrum for $D=3, ga^2=0.05$ and $K=1000$ is shown in Fig.~\ref{fig:mdep}. All eigenvalues are purely real, and the lowest-lying eigenvalue $e_0$ corresponds to the results shown in Fig.~\ref{fig:two}. The first excited state $e_1$ is distinct from the other eigenvalues for all values of $m_B^2 a^2$ shown, while the higher lying states $e_2,e_3,e_4$ are degenerate or almost degenerate.

Let me compare this to the energy spectrum of a purely quadratic field theory defined by (\ref{mainaction}) with $g=0$ and fixed particle mass $m_B=M$. In this case, using a Fourier transform in the spatial coordinate, the partition function $Z={\rm Tr}e^{-\beta {\cal H}}$ has the known expression \cite{Laine:2016hma}
\be
\label{scalarZ}
Z=\prod_k \frac{1}{2 \sinh\left(\frac{\beta \sqrt{k^2+M^2}}{2}\right)}=e^{-\sum_k \frac{\beta \sqrt{k^2+M^2}}{2} +\ln\left[1-\exp{\left(\beta \sqrt{k^2+M^2}\right)}\right]}\,,
  \ee
  where the product/sum is over all Fourier momenta $k$ on the lattice. For small temperatures $\beta\gg 1$, one can expand (\ref{scalarZ}) as
  \be
  Z=e^{-\beta E_0}+e^{-\beta E_1}+e^{-\beta E_2}+\ldots\,,
  \ee
  and read off the corresponding spectrum of the free theory as
  \be
  \label{comparisonmmm}
  E_0=\sum_k \frac{\beta \sqrt{k^2+M^2}}{2}\,,\quad
  E_1=E_0+M\,,\quad
  E_2=E_0+\sqrt{M^2+k_{\rm min}^2}\,,
  \ee
  where $k_{\rm min}$ is the minimum momentum on the lattice. If there are multiple momenta on the lattice which have the same absolute value, then $E_2$ will be degenerate with other energy values.

Comparison with the Hamiltonian spectrum shown in Fig.~\ref{fig:mdep} then suggests that $e_0$ is the vacuum energy of the system, the difference $e_1-e_0$ is the mass $M$ of the lightest excitation in the system, and $e_2,e_3,e_4$ correspond to energy states of the excitation with mass $M$ moving with lattice momentum $k_{\rm min}$.
  
 \subsection{Expectation values}

\begin{figure}[t]
  \includegraphics[width=.8\linewidth]{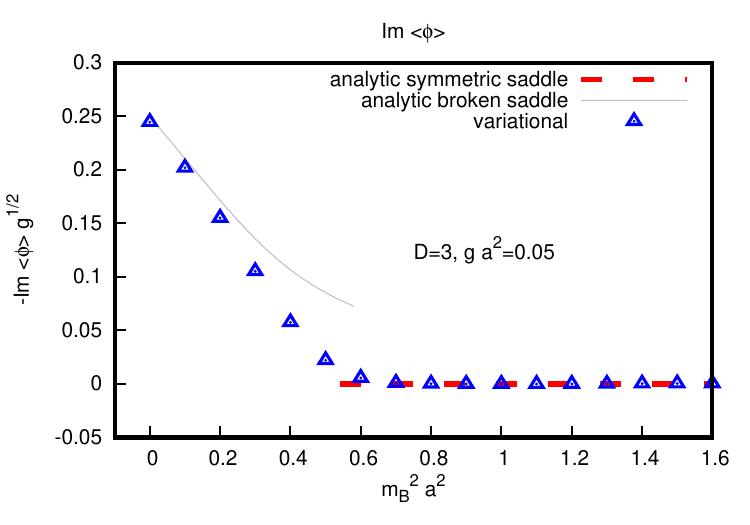}
  \caption{\label{fig:three} Field expectation value $\langle \phi\rangle$ for negative coupling $\phi^4$ theory. See text for details.}
\end{figure}
 
 Using the variational method, one can calculate the following quantities for the non-Hermitian quantum field theory:
 \begin{itemize}
 \item
   The energy spectrum $e_n$ of the quantum field theory as eigenvalues of the matrix $M_H$:
   \be
   \label{eigenvaluesetup}
    \langle n_1 n_2\ldots n_D|{\cal H}|m_1 m_2\ldots m_D\rangle \chi^{(n)}= e_n \chi^{(n)}
    \ee
  \item
    From the energy spectrum, the correlation length $C_L$ as the inverse mass gap of the system:
    \be
    \label{clm1}
  C_L \equiv \frac{1}{e_1-e_0}
  \ee
\item
  Alternatively, the correlation length $C_L$ can be obtained through the second derivative of the ground state energy, cf. Eq.~(46) \cite{Romatschke:2026qxm}
  \be
  \label{clm2}
  C_L^2=\frac{1}{D a}\frac{\partial^2 e_0}{\partial^2  m_B^2}\,.
  \ee
  I will show that at least in some parameter regime both (\ref{clm1}) and (\ref{clm2}) give identical results for $C_L$.
 \item
   Expectation values such as $\langle \phi\rangle$ in the ground state of the theory as
   \be
   \langle \phi\rangle = \frac{1}{D}\frac{\langle \tilde \chi^{(0)}| \vec{x} | \chi^{(0)}\rangle}{\langle \tilde \chi^{(0)}| \chi^{(0)}\rangle}\,. 
   \ee
 \end{itemize}

 A representative plot of the field expectation value $\langle \phi\rangle$ is shown in Fig. ~\ref{fig:three} for the case of $D=3, g a^2=0.05$. Depending on the bare mass parameter $m_B^2$, one can clearly identify two different regimes, one where $\langle \phi\rangle\neq 0$ and another where the expectation value is consistent with zero. The approximate analytic saddle-point expansion result \cite{Romatschke:2026zvd} is shown in Fig.~\ref{fig:three} for comparison, where $\langle \phi \rangle =0$ for the symmetric saddle, but $\langle \phi \rangle = \phi_0 = - \frac{i \tilde M}{8 g}\neq 0 $ from Eq.~(\ref{phi0SP}) for the broken phase saddle\footnote{One should emphasize that the field expectation value is purely imaginary, which does not happen in Hermitian quantum field theories. This is consistent with the expectations from ${\cal PT}$ symmetric quantum mechanics, where $\langle x\rangle$ also is imaginary cf.~\cite{Bender:2023cem}. The authors of Ref.~\cite{Bender:2023cem} point out that in ordinary quantum mechanics, $\langle x \rangle$ is real because $x$ is a Hermitian operator, and therefore its expectation value can be associated with the observable position of a particle. However, in ${\cal PT}$-symmetric quantum mechanics, the operator $x$ is not invariant under ${\cal PT}$ transformations, and hence its expectation value is not a observable.}. The saddle point approximation suggests a first order phase transition from broken to symmetric phase with a corresponding jump in the value of $\langle \phi\rangle$.

 By contrast, while the results from the variational method indicate a qualitative change of $\langle \phi\rangle$ around $m_B^2 a^2\simeq 0.6$, Fig. ~\ref{fig:three} does not indicate a discontinuous jump of $\langle \phi\rangle$.

\begin{figure}[t]
  \includegraphics[width=.8\linewidth]{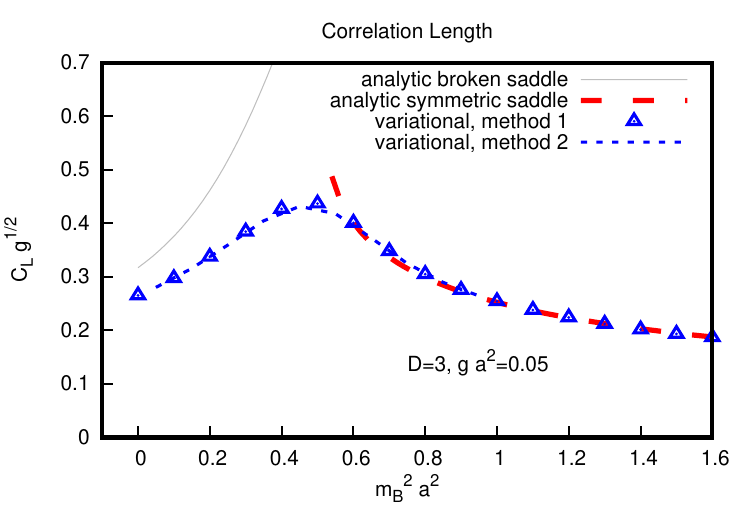}
  \caption{\label{fig:four} Correlation length $C_L$ for negative coupling $\phi^4$ theory. See text for details.}
\end{figure}

 A representative plot of the correlation length $C_L$ is shown in Fig. ~\ref{fig:four} for the case of $D=3, g a^2=0.05$. Two methods for calculating $C_L$ from Eq.~(\ref{clm1}) and Eq.~(\ref{clm2}) are shown for the variational method in the figure, and one observes that both methods agree for the parameter regime covered in the figure. In addition, results from the approximate analytic saddle-point expansion result \cite{Romatschke:2026zvd} are shown in Fig.~\ref{fig:three} for comparison, where the correlation length is defined as the inverse of the mass parameter $\tilde M, M$ for the broken and symmetric saddle, respectively, cf. appendix \ref{sec:SP}.

 The results from the variational method for $C_L$ indicate a broad peak located at $m_B^2 a^2\simeq 0.5$. The peak location for $C_L$ in Fig.~\ref{fig:four} therefore differs from the onset location of a non-vanishing $\langle \phi\rangle$ observed in Fig.~\ref{fig:three}. Taken together, these results suggest the present of a broad cross-over transition for the negative coupling theory in this coarse lattice representation.

 \subsection{Volume Dependence}

 \begin{figure}[t]
  \includegraphics[width=.8\linewidth]{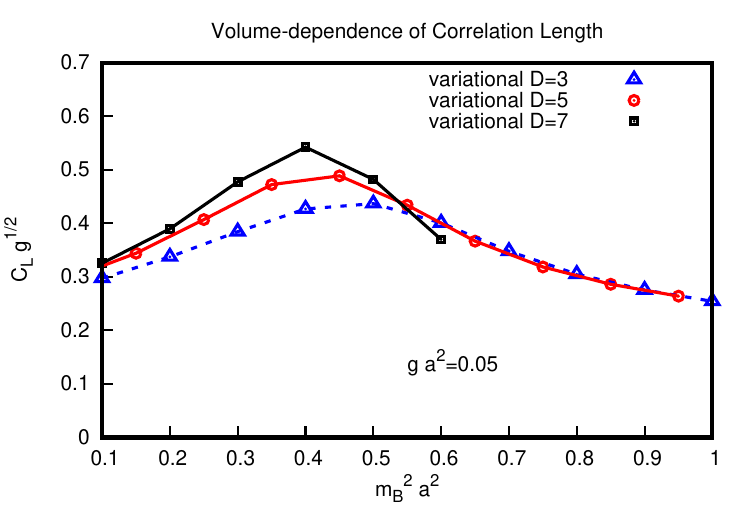}
  \caption{\label{fig:five} Volume dependence of correlation length $C_L$ for negative coupling $\phi^4$ theory. See text for details.}
 \end{figure}
 
 The variational method can be extended to larger volumes by increasing the number of transverse lattice sites $D$. However, I find that increasing $D$ requires an exponential increase in the number of basis states $K$. Specifically, while $K=10^3$ is sufficient for numerically accurate results for $D=3$, that number increases to $K=10^4$ for $D=5$ and $K=10^5$ for $D=7$. This severely limits the method to a small number of lattice sites.
  
 However, robust numerical results can be obtained for observables in non-Hermitian field theory up to at least $D=7$. This can serve multiple purposes, such as providing a benchmark for testing novel algorithms or quantum computing approaches. In addition, the results up to $D=7$ may also provide qualitative insights into the properties of non-Hermitian field theories.

 For example, the volume dependence of the correlation length $C_L$ for the negative coupling quartic theory is shown in Fig.\ref{fig:five}. From this figure, one observes that the peak in the correlation length narrows and increases in height as the volume is increased at fixed $g a^2=0.05$. The available volumes are insufficient to perform a large volume extrapolation, but the peak in $C_L$ can be used to define a transition location in the bare parameter $m_B^2 a^2$ for all volumes. In addition, the decreasing peak width with increasing volume is indicative of a strengthening transition or cross-over.

 \begin{figure}[t]
  \includegraphics[width=.8\linewidth]{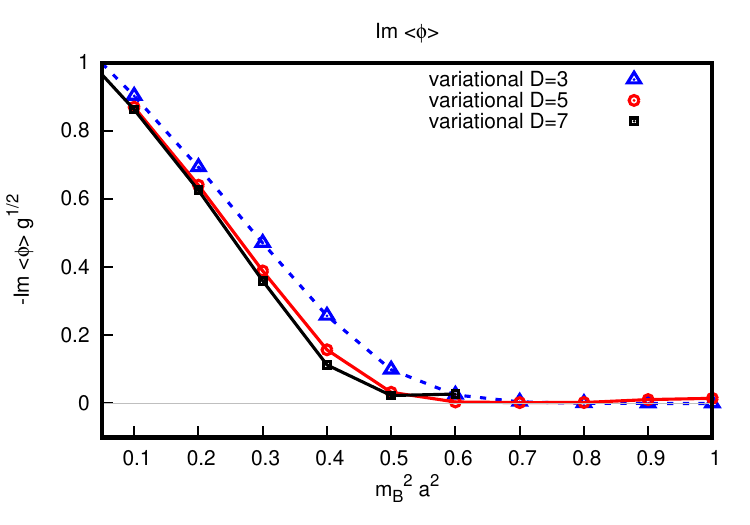}
  \caption{\label{fig:six} Volume dependence of $\langle \phi\rangle$ for negative coupling $\phi^4$ theory. See text for details.}
 \end{figure}

A similar trend is found for the volume dependence of the field expectation value $\langle \phi\rangle$ shown in Fig.~\ref{fig:six}. The transition from vanishing expectation value for large values of $m_B^2a^2$ to $\langle \phi\rangle \neq 0$ is robust up to $D=7$, and the transition is found to become more rapid as the volume is increased. 

\subsection{Finite Temperature}

Using the energy spectrum $e_n$, it is possible to consider properties of the theory at finite temperature $T=\frac{1}{\beta}$. Most straightforwardly, the Hamiltonian spectrum may be used to calculate the finite-temperature pressure $p(T)$ and the entropy density $s(T)$ of the system as
\be
\label{pfromspectrum}
p(T)=\frac{T}{L}\ln\left[\sum_{n=0}^K e^{-\beta e_n^{(K)}}\right]\,,
  \quad
  s(T)=\frac{1}{L}\ln\left[\sum_{n=0}^K e^{-\beta e_n^{(K)}}\right]+\frac{\sum_{n=0}^K e^{-\beta e_n^{(K)}} e_n^{(K)}}{L T \sum_{n=0}^K e^{-\beta e_n^{(K)}}}\,,
  \ee
  where $L=D a$ can be used to evaluate the above expressions in lattice units.

  Alternatively, one may use a symmetry of the above theory in Euclidean field theory formulation: the calculation of the Hamiltonian spectrum at zero temperature $\beta=\infty$ and finite lattice spacing $L=D a$ is equivalent to the spectrum at infinite volume $L=\infty$ and finite temperature $\beta=D a$ when relabeling space and time directions. As a consequence, the pressure for the system at infinite volume and finite temperature is given in lattice units by
  \be
  \label{pfromdual}
  p\left(T=\frac{1}{D a}\right) a^2=-\frac{e_0^{(K)} a}{D}\,,
  \ee
  which can be evaluated for fixed $m_B^2 a^2, g a^2$ using the spectral results for fixed D. For example, for the case $m_B^2a^2=0.5$ and $g a^2=0.05$ one finds the results shown in Tab.\ref{tabi}.

  \begin{table}[b]
    \begin{tabular}{|c|c|c|c|c|c|c|}
      \hline
      D & 2 & 3 & 4 & 5 & 6 & 7\\
      \hline
      $T/\sqrt{g}$ & 2.236 & 1.491 & 1.118 & 0.894 & 0.745 & 0.639\\
      \hline
      $e_0 a$ & 1.339 & 2.122 & 2.877 & 3.617 & 4.351 & 5.08\\
      \hline
      -Im $\langle \phi\rangle/a $ & 0.627 & 0.099 & 0.053 & 0.033 & 0.024 & 0.02\\
      \hline
    \end{tabular}
    \caption{\label{tabi} Table of lowest-lying Hamiltonian eigenvalue $e_0$ and field expectation value $\langle \phi\rangle$ in the variational formulation for $m_B^2 a^2=0.5$, $g a^2=0.05$ when varying $D$.}
  \end{table}

\begin{figure}[t]
  \includegraphics[width=.8\linewidth]{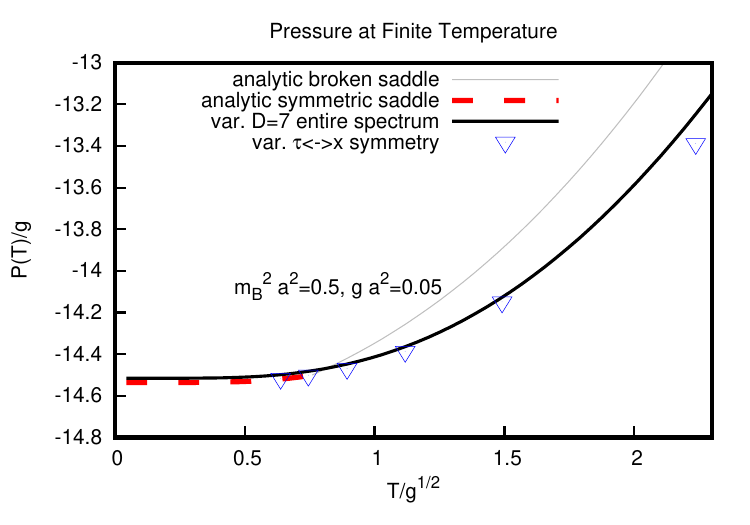}
  \caption{\label{fig:finiteT} Pressure as a function of temperature for $m_B^2 a^2=0.5, g a^2=0.05$. Shown are variational method results from the partition function (\ref{pfromspectrum}) using the entire Hamiltonian spectrum for $D=7$ (thick line) and the space-time symmetry results (\ref{pfromdual}) using only the ground state energy from $D=2$ to $D=7$ (symbols), see Tab.~\ref{tabi}. In addition, results from the approximate saddle point method (\ref{pfromsaddlem}) are shown (thin and dashed line, respectively). See text for details.}
\end{figure}

A plot of the finite temperature pressure for the case $m_B^2 a^2=0.5$, $g a^2=0.05$ is shown in Fig.~\ref{fig:finiteT}. In this figure, results from the variational method using the entire Hamiltonian spectrum for $D=7$ using (\ref{pfromspectrum}) are compared to the results obtained from the space-time symmetry, cf. Eq.~(\ref{pfromdual}) and Tab.~\ref{tabi}, respectively. The results from the variational method are compared to the results from the approximate saddle-point method (\ref{pfromsaddlem}) for $D=7$. The saddle-point method predicts a transition from symmetric saddle (at low temperature) to broken saddle (at high temperature). Both results from the variational method are consistent with each other, as well as the symmetric phase saddle result at low temperature, and differ from the broken phase saddle-point result at higher temperature.

\begin{figure}[t]
  \includegraphics[width=.8\linewidth]{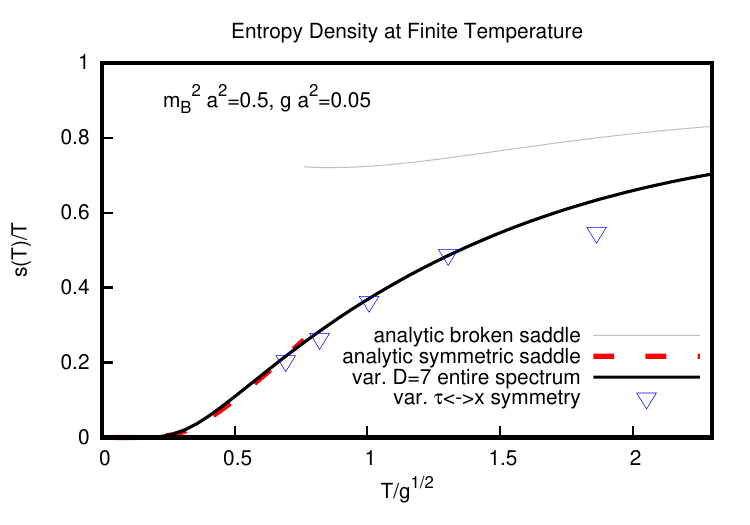}
  \caption{\label{fig:entropy} Entropy density as a function of temperature for $m_B^2 a^2=0.5, g a^2=0.05$. Shown are variational method results from the partition function (\ref{pfromspectrum}) using the entire Hamiltonian spectrum for $D=7$ (thick line) and the space-time symmetry results (\ref{pfromdual}) using only the ground state energy from $D=2$ to $D=7$ (symbols), see table \ref{tabi}. In addition, results from the approximate saddle point method (\ref{sfromsaddle}) are shown (thin and dashed line, respectively). See text for details.}
\end{figure}

A similar behavior is encountered for the entropy density shown in Fig.~\ref{fig:entropy}. The variational result for the entropy density using the entire spectrum for $D=7$ is readily calculated using (\ref{pfromspectrum}), whereas to obtain the entropy density from the space-time symmetry, Eq.~ (\ref{pfromdual}), finite differencing of the $e_0$ values in Tab.~\ref{tabi} was used. The approximate saddle-point method suggests a jump in the entropy density around $T_c\simeq 0.7 \sqrt{g}$, which is not observed using the results from the variational method. This should be contrasted with the observed turn-on of the field expectation value $\langle \phi\rangle$ for high temperatures reported in Tab.~\ref{tabi}.

\begin{figure}[t]
  \includegraphics[width=.8\linewidth]{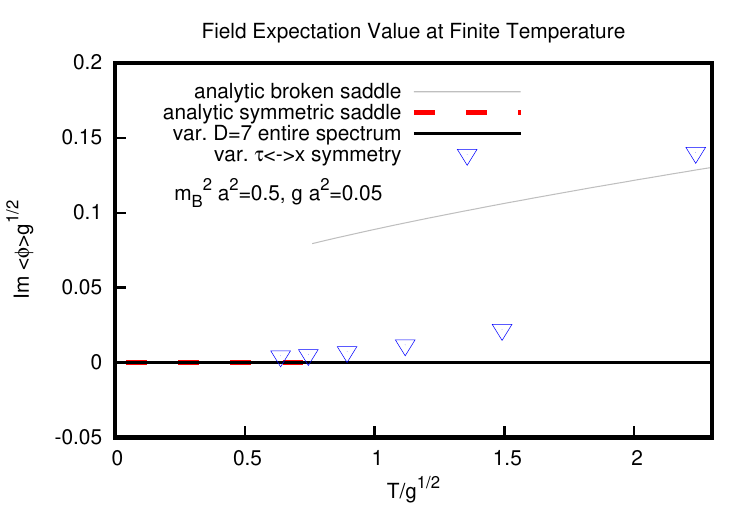}
  \caption{\label{fig:phivalT} Field expectation value as a function of temperature for $m_B^2 a^2=0.5, g a^2=0.05$. Shown are variational method results from the partition function (\ref{pfromspectrum}) using the entire Hamiltonian spectrum for $D=7$ (thick line) and the space-time symmetry results (\ref{pfromdual}) using only the ground state energy from $D=2$ to $D=7$ (symbols), see Tab.~\ref{tabi}. In addition, results from the approximate saddle point method (\ref{phi0SP}) are shown (thin and dashed line, respectively). See text for details.}
\end{figure}

Results for ${\rm Im}\langle \phi\rangle$ are shown as a function of temperature in Fig.~\ref{fig:phivalT} for $m_B^2 a^2=0.5, g a^2=0.05$. Since the variational result for $D=7$ implies vanishingly small expectation value, $\langle\phi\rangle=0$ has been used for the Hamiltonian spectrum result in this figure. By contrast, the space-time symmetry results for $\langle \phi\rangle$ from Tab.~\ref{tabi} imply a non-vanishing result at higher temperatures. For comparison, the results from the approximate saddle-point method (\ref{phi0SP}) are also shown. 

Taken as a whole, the lattice record for the finite-temperature properties with lattice parameters $m_B^2 a^2=0.5, g a^2=0.05$ remains inconclusive. Pressure and entropy density in the variational method do not exhibit signs of a phase transition or rapid cross-over, even though the field expectation value $\langle \phi\rangle$ does indicate a gradual change in behavior. Higher resolution on larger lattices would be a prerequisite to study the situation further.

\subsection{Road-map ahead: exponential reduction of states}

As outlined above, the variational method is capable of precision calculations for non-Hermitian field theories in the Hamiltonian formulation in lattice discretization. However, the method as presented suffers from the exponential cost increase stemming from the required number of basis states increasing exponentially with the number of lattice sites $D$.

For the case of negative coupling $\phi^4$ theory, for the lattice parameters studied in this work, it was found that the ground-state wave-function remains symmetric under ${\cal PT}$-symmetry. Since ${\cal PT}$-symmetry is not broken in the ground state, this suggests a potential way forward for reducing the effective number of basis states for the ``static'' eigenvectors of the Hamiltonian, e.g. those that have vanishing lattice momentum $k$, cf. Eq.~(\ref{comparisonmmm}). Since those vectors correspond to zero momentum excitations, the basis elements (\ref{basiselement}) must be symmetric in the spatial coordinates $x_1,x_2,\ldots x_D$. Put differently, the eigenvectors $\chi^{(n)}$ as the solution to (\ref{eigenvaluesetup}) corresponding to zero-momentum excitations must have the same weights for all basis states $|m_1 m_2\ldots m_D\rangle$ where $m_1,m_2,\ldots,m_D$ are permutations of each other. I have verified this using the above variational method, and find that both the ground state vector $\chi^{(0)}$ as well as the first excited state vector $\chi^{(1)}$ obey this rule.

As a consequence, these effective zero-momentum states may be constructed more simply as
\be
\langle \vec{x}| n_1 n_2\ldots n_D \rangle_{k=0} = \frac{1}{\sqrt{\# perm.}}\left(\psi_{n_1}(x_0)\psi_{n_2}(x_1)\ldots \psi_{n_D}(x_{D-1})+{\rm perm.}\right)\,,
\ee
with the normalizing prefactor given by the number of permutations. Since the number of permutations grows exponentially with the number of lattice sites $D$, the zero-momentum states capture the rapidly increasing number of states discussed above while only counting as a ``single'' state in the effective zero-momentum basis. The price to pay for the reduced number of states is that instead of the matrix elements (\ref{QFTmatrixelements}) needed in (\ref{eigenvaluesetup}) one now needs the sum over permutations
\be
\sum_{{\rm perm}(n_i),{\rm perm(m_i)}} I_{n_1n_2\ldots n_D}^{m_1 m_2\ldots m_D}\,,
  \ee
  to evaluate a single zero-momentum state matrix element. Despite this cost, the dramatic reduction in the size of the matrix for the eigenvalue problem suggests that using this method, lattice sizes in excess of ${\cal O}(10)$ will be accessible.

\section{Summary and Conclusions}
\label{sec:conc}

In this work, I implemented a variational method to study non-Hermitian quantum field theory in two dimensions in the Hamiltonian formulation. It was found that the standard implementation of the variational method needs to be adapted in order to treat non-Hermitian systems, and I pointed out how this can be achieved in practice for the example of negative coupling quartic self-interaction. I presented results of the method for fixed lattice spacing and small lattice volumes to explore the parameter space of the theory, finding two regimes which differ by the order parameter $\langle \phi\rangle$, the expectation value of the field. The results point to a cross-over transition on the lattice between the two regimes, which is present for all volumes and appears to strengthen as the volume is increased.

Furthermore, properties of the field theory at finite temperature were investigated.

No spontaneous breaking of ${\cal PT}$-symmetry for any of the lattice parameters investigated is observed in this work. Instead, I found a purely real energy eigenspectrum whenever employing a sufficiently high number of variational basis states $K$.

This requirement also limits the applicability of the method to large volumes, as the number of states $K$ required to obtain stable energy eigenspectra appears to be growing exponentially in volume. Nevertheless, the variational method discussed in this work performs superior to brute-force numerical integration of non-Hermitian theories on the lattice \cite{Romatschke:2023fax}, which is limited to even smaller volumes. With presently available computing resources, lattice volumes of up to ${\cal O}(10)$ are feasible, which could be enough to explore qualitative features of the phase diagram of non-Hermitian field theories in d=2. In addition, I have reported on a potential refinement of the method that reduces the number of required basis states, which should make variational calculations of ground-state properties possible for volumes in excess of ${\cal O}(10)$. 

Comparison of the variational method to the semi-analytic saddle-point approximations outlined in Ref.~\cite{Romatschke:2026zvd} presented in this work corroborate the finding of Ref.~\cite{Romatschke:2026qxm} that the approximate method correctly represents regions of interest of the phase diagram, such as those where the order parameter changes.

Together, these points suggest that the phase diagram of non-Hermitian field theory can be explored efficiently by first using the approximate saddle point methods to narrow the lattice parameter space to be subsequently explored by the variational method. I hope to report on this approach in the future.

  \section{Acknowledgments}

  I would like to thank A. Miscioscia for helpful discussions.  Die präsentierten Rechenergebnisse wurden zum Teil am Vienna Scientific Cluster (VSC) erzielt.

 \begin{appendix}
   \section{Saddle-Point Method}
   \label{sec:SP}

   In an effort to keep this work self-contained, this appendix outlines the saddle-point method introduced and studied in Refs.~\cite{Romatschke:2026tam,Romatschke:2026qxm}.

   The saddle-point method calculates approximate results for the partition function $Z=\int {\cal D}\phi e^{-S}$ with $S$ given by the Euclidean action (\ref{mainaction}) or alternatively its discretized version (\ref{latticeaction}). Two different saddle points are used, one built around a symmetric configuration and one where the action is expanded around $\phi=\phi_0={\rm const.}$.

   The detailed derivation of the approximate partition function for the symmetric and broken saddle, respectively, was the subject of Ref.~\cite{Romatschke:2026tam}, and the interested reader is referred to this article. The result of repeating the steps from Ref.~\cite{Romatschke:2026tam} for the theory defined by the action (\ref{latticeaction}) can be summarized as follows:
   \begin{itemize}
   \item
     For the symmetric-phase saddle, the local pressure is given by
     \be
     p(M)=p_{\rm free}(M)-\frac{(M^2-m_B^2)^2}{48g}\,,\quad M^2=m_B^2-12 g G_{\rm free}(M)\,,
     \ee
     where $M$ is the mass of the propagating excitation in the symmetric phase.
   \item
     For the broken-phase saddle, the local pressure is given by
     \be
     \tilde p(\tilde M)=p_{\rm free}(\tilde M)+\frac{\tilde M^4}{96 g}+\frac{\tilde M^2 m_B^2}{24 g}-\frac{m_B^4}{48g}\,,\quad -\frac{\tilde M^2}{2}=m_B^2-12 g G_{\rm free}(\tilde M)\,,
     \ee
     where $\tilde M$ is the mass of the propagating excitation in the broken phase that is related to the constant field value $\phi_0$ as
     $\phi_0^2=-\frac{\tilde M^2}{8 g}$, cf. Ref.~\cite{Romatschke:2026zvd}. Unlike the case of the positive coupling theory, for $\lambda=-g<0$ the Lefshetz thimble corresponding to the saddle has non-vanishing intersection number only for
     \be
     \label{phi0SP}
     \phi_0=-\frac{i \tilde M}{\sqrt{8g}}\,,
     \ee
     so that $\langle \phi\rangle=\phi_0\neq 0$ for the broken phase saddle.
   \item
     In the above formulas, $p_{\rm free}(m),G_{\rm free}(m)$ refer to the pressure and two-point field correlation function of a free particle with mass $m$. In the zero temperature limit, one has
     \ba
     \label{pfreedef}
     p_{\rm free}(m)&=&-\frac{1}{2L}\sum_n \int \frac{d\omega}{2\pi}\ln\left[\omega^2+m^2+\Omega_n^2(D)\right]=-\frac{1}{2L}\sum_n \sqrt{m^2+\Omega_n^2(D)}\,,\\
     G_{\rm free}(m)&=&-2 \frac{\partial p_{\rm free}(m)}{\partial m^2}=\frac{1}{2L}\sum_n \frac{1}{\sqrt{m^2+\Omega_n^2(D)}}\,,
     \ea
     where $\Omega_n^2(D)=\frac{4}{a^2}\sin^2\left(\frac{\pi n}{D}\right)$ is the lattice Matsubara frequency squared.
   \item
     Given fixed bare parameters $m_B^2 a^2,g a^2$ and a fixed number of spatial lattice sites $D$, one evaluates the excitation masses $M a, \tilde M a$ and partial pressures $a^2 p(M) ,a^2\tilde p(\tilde M)$ for the symmetric and broken phase saddle, respectively. In the zero temperature limit one can relate the pressure to the ground-state energy $E_0$ of the Hamiltonian as
     \be
     \label{relationSP}
     E_0a =-L p a=-D a^2 p\,,
     \ee
     which follows from $Z=e^{\beta L p}\simeq e^{-\beta E_0}$.
   \item
     At non-zero temperature $T=\frac{1}{\beta}\neq 0$, the frequency integral (\ref{pfreedef}) turns into a sum over Matsubara frequencies $\omega_n=2 \pi n T$ with $n\in \mathbb{N}$, so that using the short-hand notation $\epsilon_n\equiv \sqrt{m^2+\Omega_n^2(D)}$ one has
     \ba
     \label{pfromsaddlem}
     p_{\rm free}(m,T)&=&-\frac{1}{2L}\sum_n \epsilon_n-\frac{T}{L}\sum_n \ln\left(1-e^{-\beta \epsilon_n}\right)\,,\nonumber\\
     G_{\rm free}(m,T)&=&\frac{1}{2L}\sum_n \frac{1+2 n_B\left(\beta \epsilon_n\right)}{\epsilon_n}\,,
     \ea
     where $n_B(x)=\frac{1}{e^{x}-1}$. Eq.~(\ref{pfromsaddlem}) can be used to calculate the partial pressures for the symmetric and broken phase at finite temperature.
   \item
     The stationarity of the saddle-points directly leads to the result for the entropy density at finite temperature
     \be
     \label{sfromsaddle}
     s(m,T)=\frac{T}{L}\sum_n \left[\epsilon_n n_B(\beta \epsilon_n)-T \ln\left(1-e^{-\beta \epsilon_n}\right)\right]\,,
     \ee
     where $m=M$ and $m=\tilde M$ for the symmetric and broken phase saddle, respectively.
       \end{itemize}

   \end{appendix}

\bibliography{enormous,enormous2}
\end{document}